\documentclass[aps,pra,twocolumn,superscriptaddress,showpacs,amsmath,amssymb]{revtex4-1}

\usepackage{braket}
\usepackage[per=slash,
            emulate=units, 
            range-phrase = -,
            range-units = single,
            exponent-product = \cdot,
            inter-unit-product = \cdot,
            separate-uncertainty = true]{siunitx} 

\usepackage{hyperref}
\usepackage{graphicx}

\begin{document}

\title{Ionization spectra of highly Stark-shifted rubidium Rydberg states}

\author{Jens Grimmel}
\email[]{jens.grimmel@uni-tuebingen.de} 
\affiliation{Center for Quantum Science, Physikalisches Institut, Eberhard-Karls-Universit\"at T\"ubingen, Auf der Morgenstelle 14, D-72076 T\"ubingen, Germany} 

\author{Markus Stecker}
\email[]{markus.stecker@uni-tuebingen.de} 
\affiliation{Center for Quantum Science, Physikalisches Institut, Eberhard-Karls-Universit\"at T\"ubingen, Auf der Morgenstelle 14, D-72076 T\"ubingen, Germany} 

\author{Manuel Kaiser}
\affiliation{Center for Quantum Science, Physikalisches Institut, Eberhard-Karls-Universit\"at T\"ubingen, Auf der Morgenstelle 14, D-72076 T\"ubingen, Germany} 

\author{Florian Karlewski} 
\affiliation{Center for Quantum Science, Physikalisches Institut, Eberhard-Karls-Universit\"at T\"ubingen, Auf der Morgenstelle 14, D-72076 T\"ubingen, Germany} 

\author{Lara Torralbo-Campo}
\affiliation{Center for Quantum Science, Physikalisches Institut, Eberhard-Karls-Universit\"at T\"ubingen, Auf der Morgenstelle 14, D-72076 T\"ubingen, Germany} 

\author{Andreas G\"unther}
\affiliation{Center for Quantum Science, Physikalisches Institut, Eberhard-Karls-Universit\"at T\"ubingen, Auf der Morgenstelle 14, D-72076 T\"ubingen, Germany} 

\author{J\'{o}zsef Fort\'{a}gh}
\email[]{fortagh@uni-tuebingen.de}
\affiliation{Center for Quantum Science, Physikalisches Institut, Eberhard-Karls-Universit\"at T\"ubingen, Auf der Morgenstelle 14, D-72076 T\"ubingen, Germany} 

\date{\today}

\begin{abstract}
	We report on the observation and numerical calculation of ionization spectra of highly Stark-shifted Rydberg states of rubidium beyond the classical ionization threshold. In the numerical calculations, a complex absorbing potential (CAP) allows us to predict the energy levels and ionization rates of Rydberg states in this regime. Our approach of adjusting the CAP to the external electric field reduces the number of free parameters from one per resonance to a single one. Furthermore, we have measured the ionization spectra of magneto-optically trapped rubidium atoms which are excited to principal quantum numbers of 43 and 70 at various electric fields. The emerging ions are detected using an ion optics. We find good agreement between the numerically and experimentally obtained spectra. 
\end{abstract}

\maketitle

\section{\label{sec:intro}Introduction}

Stark spectra of alkali-metal Rydberg states above the classical ionization threshold exhibit an intricate energy-level structure with strongly varying ionization behavior, including rapidly ionizing states as well as the extreme of narrow resonances where ionization is almost suppressed \cite{Gallagher.1994}. The latter feature is clearly distinct from hydrogen, where the ionization rate of a given state grows exponentially with the applied external electric field \cite{Damburg.1979}. The study of these ionization spectra of non-hydrogenic atoms is therefore of particular interest not only from a fundamental point of vie,w but also for the prospect of improved control over the ionization process in Rydberg gases. 

The method of complex rotation (CR) is well known for the theoretical treatment of ionization rates of highly Stark-shifted states of hydrogen and alkali atoms \cite{Reinhardt.1976,Reinhardt.1982,Stevens.1996}. It is applied by substituting the location and momentum operators in the Hamiltonian by the complex terms $\hat{r}\rightarrow \hat{r}\cdot\exp(i\theta)$ and $\hat{p}\rightarrow \hat{p}\cdot\exp(-i\theta)$, respectively, which results in a non-Hermitian Hamiltonian. This leads to complex eigenvalues of the Hamiltonian which can be used to obtain the energy levels and linewidths, i.e., the ionization rates in the present case. As an alternative to this method, a complex absorbing potential (CAP) can be employed to create a non-Hermitian Hamiltonian \cite{Kosloff.1986,Riss.1993,Sahoo.2000,Ho.2013}. The CAP is added to the original Hamiltonian in the form $-i\eta W(\vec{r})$. Both of these methods, CR and CAP, work with free parameters $\theta$ and $\eta$, respectively, which are determined for every single resonance of the system by a variational method. 

In a previous work \cite{Grimmel.2015} we have calculated Stark-shifted energy levels including the corresponding dipole matrix elements by diagonalization of a matrix representation of the Hamiltonian \cite{Zimmerman.1979}. While this methods yields very precise results for spectra at electric fields below the classical ionization threshold, the broadening of the states by ionization at higher fields can no longer be neglected. Hence, in this work, we combine the matrix diagonalization method with an adaptive CAP method by choosing a potential that is adjusted to the external electric field. This removes the need to determine the free parameter $\eta$ for each resonance separately and thereby greatly reduces the computational effort for the numerical calculation. We calculate ionization spectra near the unperturbed $43\text{S}_{1/2}$ and $70\text{S}_{1/2}$ $^{87}$Rb Rydberg states for electric fields far beyond the classical ionization threshold. Furthermore, we present an experiment in which rubidium atoms in a magneto-optical trap (MOT) are excited to Rydberg states in the presence of an external electric field. When the atoms ionize from these Stark-shifted Rydberg states, the ions are guided to a detector by an ion-optical system \cite{Stecker.2017}. We have measured ionization spectra by ramping up the external electric field and scanning the excitation laser frequency near the aforementioned states. 

The methods we present here can be used to search for resonances from the ionization spectra, such as highly Stark-shifted states which have a desirable ionization rate or sensitivity to the external electric field across a certain range. Furthermore, this opens up the possibility to tune a coupling between ionizing and non-ionizing states by the external electric field. This way of tailoring the ionization process is highly useful for the design of sources of cold ions and electrons for microscopy purposes \cite{Kime.2013,Murphy.2015,Thompson.2016,Moufarej.2017}. Moreover, a precise knowledge of these ionization spectra opens new perspectives for experiments incorporating Rydberg atoms near surfaces, where static electric fields arise due to adsorbates \cite{Tauschinsky.2010,Abel.2011,Hattermann.2012,Chan.2014}. 

\newpage

\section{\label{sec:th}Application of a complex absorbing potential}

\begin{figure}
	\includegraphics{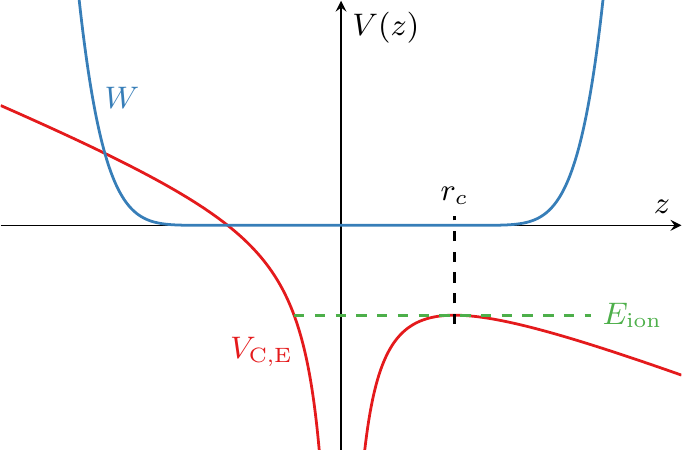}
	\caption{\label{fig:coulomb_potential_and_cap}(Color online) Illustration of the complex absorbing potential (CAP). The parameter $r_\mathrm{c}$ is determined according to Eq. \eqref{eq:rc} which places it at the radius of the saddle point of the potential $V_\mathrm{C,E}$ (see Eq. \eqref{eq:vce}). Note that in this 1D graph along the $z$ axis the saddle point appears as a local maximum. This choice of $r_\mathrm{c}$ creates a potential $W$ that resembles the shape of $r^6$ but is spherically shifted by $r_\mathrm{c}$.} 
\end{figure} 

The Hamiltonian $\hat{H}$ for an atom in an external electric field $F_\mathrm{E}$ along the $z$ axis is given by  
\begin{equation} 
	\hat{H} = \hat{H}_0 + F_\mathrm{E}\hat{z},
\end{equation} 
where $\hat{H}_0$ denotes the unperturbed Hamiltonian, i.e., the unperturbed atomic energy levels, which are calculated from quantum defect theory \cite{Li.2003,Mack.2011,Han.2006,Afrousheh.2006}. Note that all formulas in this section are given in atomic units. The Stark-shifted energy levels corresponding to this Hamiltonian are routinely calculated by choosing a subset of the basis given by $\hat{H}_0$, representing $\hat{H}$ as a matrix in this basis and computing the eigenvalues of this matrix \cite{Zimmerman.1979}. For the high-field region considered in this work, it is crucial to include all total angular momentum quantum numbers $j$ in the subset of the basis. However, as we use a two-photon excitation scheme in the experiment, we can limit our calculations to $|m_j| \in \{1/2,3/2,5/2\}$. The subset of the basis is then chosen symmetrically in energy above and below the desired energy region. The convergence of this method can be assured by increasing the subset of the basis until changes of the resulting eigenvalues are well below the experimental resolution. In our previous work, we have extended the calculations of \cite{Zimmerman.1979} and determined a measure $D$ for the transition strength in the three-level ladder scheme by calculating the dipole matrix elements between the states \cite{Grimmel.2015}. Here, we further extend these calculations by introducing a complex absorbing potential (CAP) to the model, which allows for an estimate of the ionization rates of Stark-shifted states in the regime of high electric fields beyond the classical ionization threshold. 

The CAP is added to the Hamiltonian $\hat{H}$, resulting in a new non-Hermitian Hamiltonian,
\begin{equation}
	\hat{H}_\mathrm{CAP} = \hat{H} - i\eta W(\hat{r},F_\mathrm{E}).
	\label{eq:hamiltonian_cap}
\end{equation}
In general, the free parameter $\eta\in\mathbb{R}^+$ should be adjusted for each resonance that is studied at each value of the electric field. However, in the approach we present in this work, we also adjust the function $W(\hat{r},F_\mathrm{E})$ depending on the electric field, which in effect allows us to choose the parameter $\eta$ only once for a whole region of the spectrum. 

We have chosen a CAP combining $\hat{r}^6$ as in \cite{Sahoo.2000} with the Heaviside function $\Theta$ as in \cite{Riss.1993}, similar to the rectangular-box CAP in \cite{Santra.2001}:
\begin{equation}
	W(\hat{r},F_\mathrm{E}) = \Theta(\hat{r}-r_\mathrm{c}(F_\mathrm{E})) \cdot (\hat{r}-r_\mathrm{c}(F_\mathrm{E}))^6.
\end{equation} 
This results in a spherical potential scaling as $\hat{r}^6$, but radially shifted to a radius $r_\mathrm{c}$. The matrix representation of this CAP is calculated using the same radial wave functions as for the $\hat{z}$ operator, which are obtained by integrating a parametric model potential \cite{Marinescu.1994}. It is worth noting that due to its spherical symmetry, this choice for the CAP does not introduce any coupling between different states and therefore results in a purely diagonal matrix representation. Furthermore, the Hamiltonian can still be treated separately for different values of $|m_j|$ and the basis for the matrix representation is chosen as in previous works by including enough states nearby in energy for the results to converge \cite{Zimmerman.1979,Grimmel.2015}. 

The potential $W(\hat{r},F_\mathrm{E})$ is changed along with the external electric field $F_\mathrm{E}$ via the radius $r_\mathrm{c}$. To determine $r_\mathrm{c}$, we use a Coulomb potential to approximate the atomic potential with an external electric field, 
\begin{equation}
	V_\mathrm{C,E} = -\frac{1}{r} - F_\mathrm{E}z.
	\label{eq:vce}
\end{equation}
This potential is also used for the definition of the classical ionization threshold, which is given by
\begin{equation}
	E_\mathrm{ion} = - 2 \sqrt{F_\mathrm{E}},
\end{equation}
and marks the saddle point where the potential opens towards the continuum. The location of the saddle point, which is a local maximum along the $z$ axis, is then given by
\begin{equation}
	r_\mathrm{c}(F_\mathrm{E}) = \frac{1}{\sqrt{F_\mathrm{E}}}
	\label{eq:rc}
\end{equation}
and we use this radius to place the onset of $r^6$ in the potential $W(\hat{r},F_\mathrm{E})$, as illustrated in Fig. \ref{fig:coulomb_potential_and_cap}. Graphically speaking, this allows us to distinguish between an inside and an outside region of the atom and to ensure that the CAP only absorbs the parts of the wave functions that protrude to the outside region. 

The Hamiltonian $\hat{H}_\mathrm{CAP}$ may yield complex eigenvalues which can be written as 
\begin{equation}
	E_\mathrm{c} = E_r - i \frac{\Gamma}{2},
	\label{eq:e_c}
\end{equation}
where the real part determines the energy level $E_r$ of the resonance and the imaginary part determines its ionization rate $\Gamma$. In the calculations the free parameter $\eta$ from Eq. \eqref{eq:hamiltonian_cap} is varied for exemplary values of the electric field. The first of these exemplary points is conveniently chosen near the classical ionization threshold to determine a first estimate for $\eta$ because, assuming no other previous knowledge about the spectrum, this region generally features a large number of resonances with different ionization rates. Subsequently, further points are chosen at higher electric fields to continue the variation of $\eta$ following the initial estimate. The parameter $\eta$ is then fixed to a value in the center of a region which leaves the resulting eigenvalues $E_c$ mostly unchanged. For all spectra investigated here, this region spans multiple orders of magnitude of $\eta$, which is a direct consequence of our field-dependent choice for the CAP. It is also worth noting that within this region, changes of $\eta$ affect the broadest resonances, i.e., broader than the frequency range in the experiment, more than narrow resonances. Therefore, the resulting changes do not visibly alter the spectrum. 

In order to create a diagram of the spectrum from the calculated eigenvalues in terms of intensity $I$ with respect to the transition energy $E$, we sum up Lorentzian peaks for each resonance at a given electric field: 
\begin{equation}
	I(E) = \frac{D}{\pi} \cdot \frac{\Gamma}{\Gamma^2 + (E-E_r)^2}.
\end{equation}
For the resulting resonances which would have linewidths smaller than \SI{25}{\MHz}, we set the linewidth to \SI{25}{\MHz}, which is approximately the linewidth of resonances in the region that is not dominated by ionization and also the resolution of our experimental data. Please note that ionization is the only decay channel that we consider in these calculations. All other decay channels as well as redistribution by blackbody radiation are negligible for strongly ionizing states.

\section{\label{sec:exp}Experimental setup}

\begin{figure}
	\includegraphics{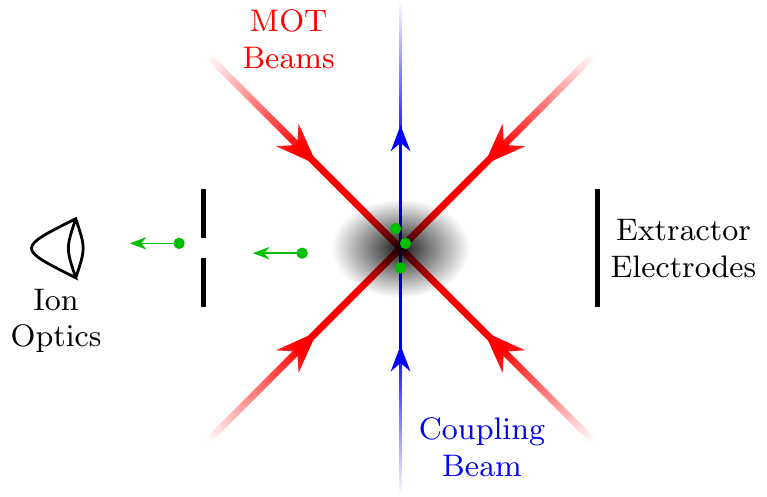}
	\caption{\label{fig:aufbau_schema_ionenoptik}(Color online) Schematic diagram of the experimental setup. Rubidium atoms are trapped at the intersection of the MOT beams (red). Atoms from the MOT are continuously excited to Rydberg states by the coupling laser (blue) at various electric fields, which are applied via the extractor electrodes. The field is also used to guide the ions (green) to the ion optics.} 
\end{figure} 

We excite Rydberg atoms in a cloud of $^{87}\text{Rb}$ atoms confined to a standard six-beam magneto-optical trap (MOT). The cooling  light is provided by a diode laser running at \SI{\approx780}{\nm}, which is frequency stabilized to the cooling transition $5\text{S}_{1/2} (F=2) \rightarrow 5\text{P}_{3/2} (F=3)$ with a red detuning of \SI{10}{\MHz}. The MOT is positioned in between two electrodes which are used both to generate the desired electric field at the position of the MOT and to extract ions out of the MOT (see Fig. \ref{fig:aufbau_schema_ionenoptik}). An ion optics consisting of a set of electrostatic lenses guides the ions to a microchannel plate detector (MCP) where they can be detected with singl- particle sensitivity \cite{Stecker.2017}. 

The excitation of rubidium atoms from the ground state to a Rydberg level is done in a two-step process. The lower transition $5\text{S}_{1/2} \rightarrow 5\text{P}_{3/2}$ is driven by the MOT beams themselves. The upper transition from the intermediate state $5\text{P}_{3/2}$ to a Rydberg level is done by a frequency-doubled, grating stabilized diode laser (Toptica DL-SHG pro) with a tuneable wavelength around \SI{480}{\nm} (hereafter referred to as ``coupling laser''). The frequency of the coupling laser is stabilized by a HighFinesse WSU-30 Wavemeter. The ionization rate of the excited Rydberg atoms is determined by counting the emerging ions with the MCP detector. 

In order to measure the ionization spectra of highly Stark-shifted Rydberg states, we fix the coupling laser to a certain frequency and ramp the voltage at the extractor electrodes to probe the desired field region and simultaneously detect the generated ions. This ramp is repeated for different coupling laser frequencies. The MOT beams and the coupling beam remain switched on during the whole measurement and the MOT is continuously loaded from rubidium dispensers. With this scheme, we were able to scan a large field region in a reasonably small measurement time, while still getting a good signal-to-noise ratio. 

We measured Stark spectra near the unperturbed $43\text{S}_{1/2}$ and $70\text{S}_{1/2}$ state in a wide electric-field range, mostly around and above the classical ionization limit. In these measurements, we used voltage ramps with a ramping speed of \SI{1}{\volt \per \second} and \SI{25}{\MHz} steps for the frequency of the coupling laser.

\section{\label{sec:results}Measurements and comparison to CAP-theory results}

\begin{figure*}
	\includegraphics{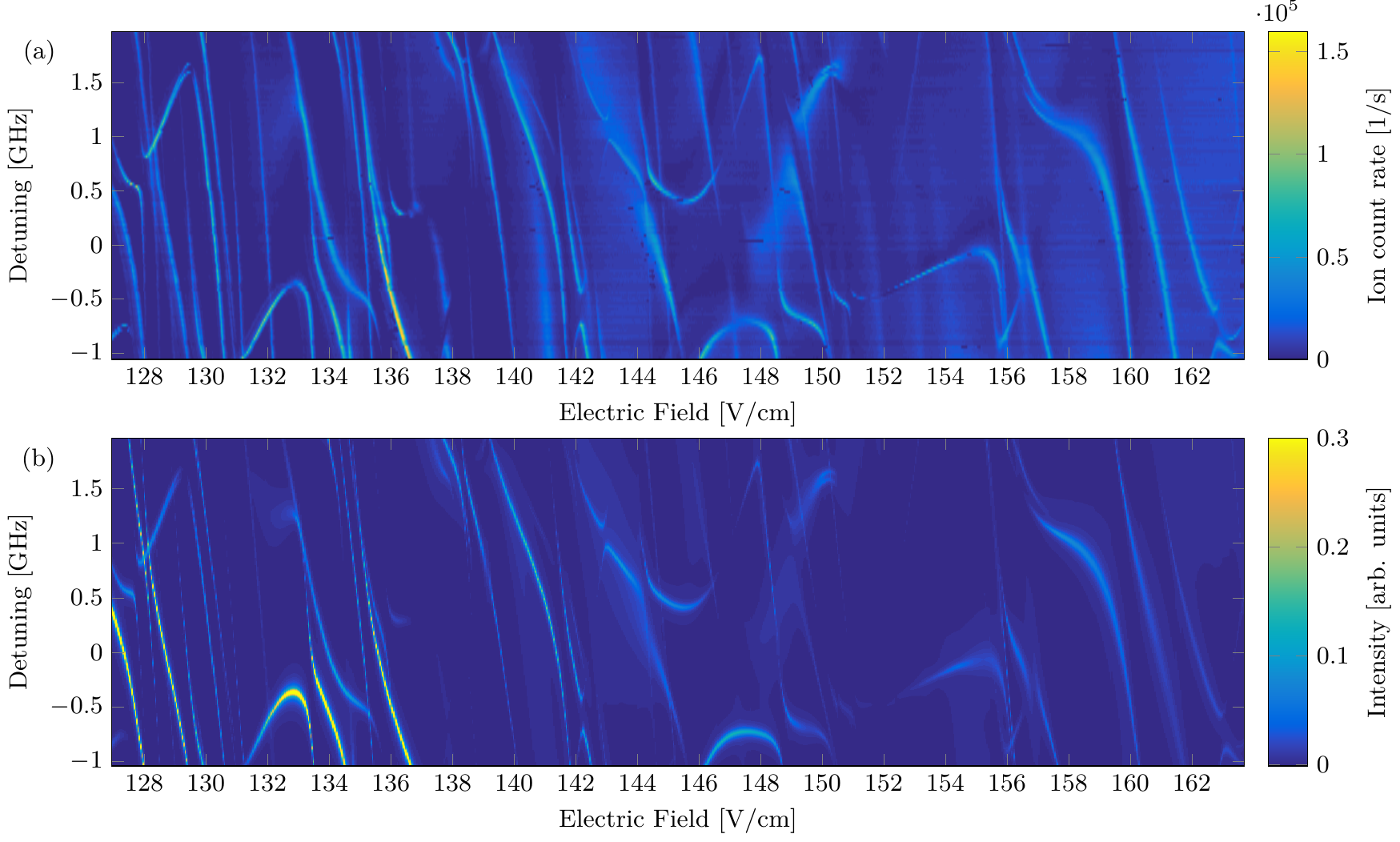}
	\caption{\label{fig:43S-1-2-exp_cap}(Color online) Ionization Stark spectrum near $43\text{S}_{1/2}$. (a) Detected ion signal from the experiment. (b) Results from the numerical calculations for $\eta=\num{1e6}$. The detuning is given relative to the unperturbed state. We have applied a linear scaling to match the electric field from the experiment to the theoretical results. The classical ionization threshold is located at \SI{\approx127}{\V\per\cm}.} 
\end{figure*} 

\begin{figure*}
	\includegraphics{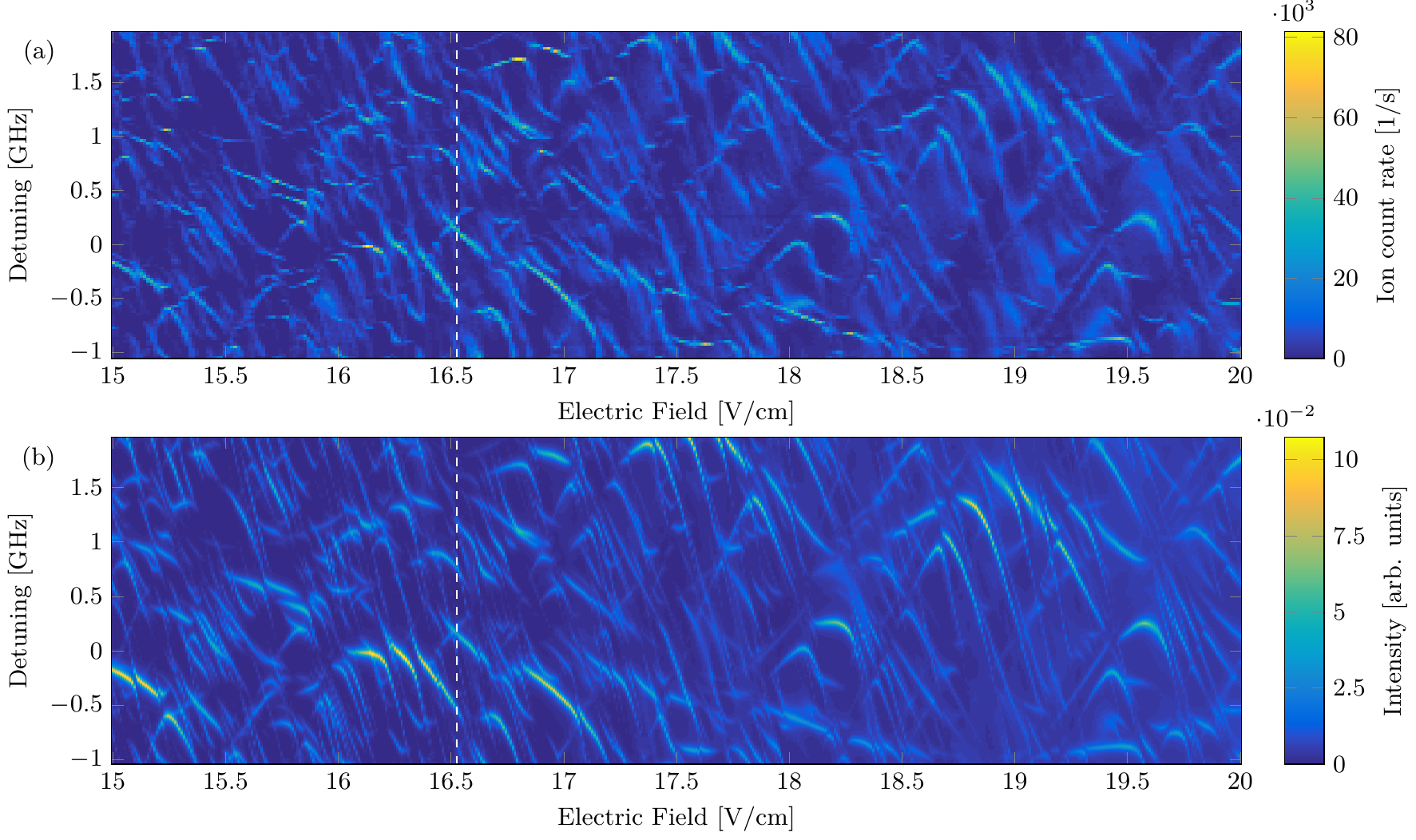}
	\caption{\label{fig:70S-1-2-exp_cap_f_ext_15_20}(Color online) Ionization Stark spectrum near $70\text{S}_{1/2}$. (a) Detected ion signal from the experiment. (b) Results from the numerical calculations for $\eta=\num{2e5}$. The detuning is given relative to the unperturbed state. We have applied a linear scaling to match the electric field from the experiment to the theoretical results. The classical ionization threshold is located at \SI{\approx16.1}{\V\per\cm}. The white dashed lines mark the cut that is shown in more detail in Fig. \ref{fig:70S-1-2_freqlines}(a).} 
\end{figure*} 

\begin{figure*}
	\includegraphics{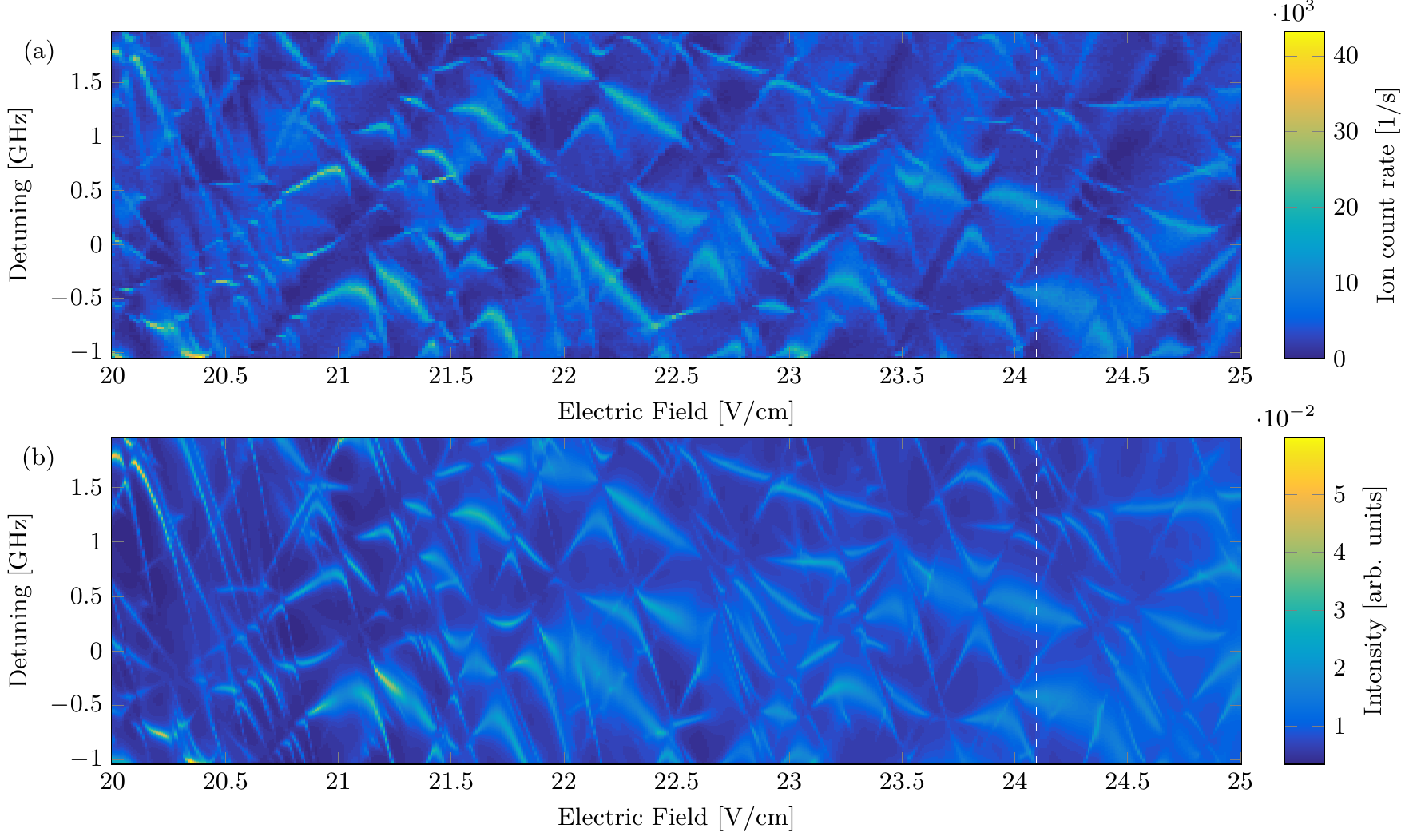}
	\caption{\label{fig:70S-1-2-exp_cap_f_ext_20_25}(Color online) Ionization Stark spectrum near $70\text{S}_{1/2}$. Continuation from Fig. \ref{fig:70S-1-2-exp_cap_f_ext_15_20}. The white dashed lines mark the cut that is shown in more detail in Fig. \ref{fig:70S-1-2_freqlines}(b).} 
\end{figure*} 
 
\begin{figure*}
	\includegraphics{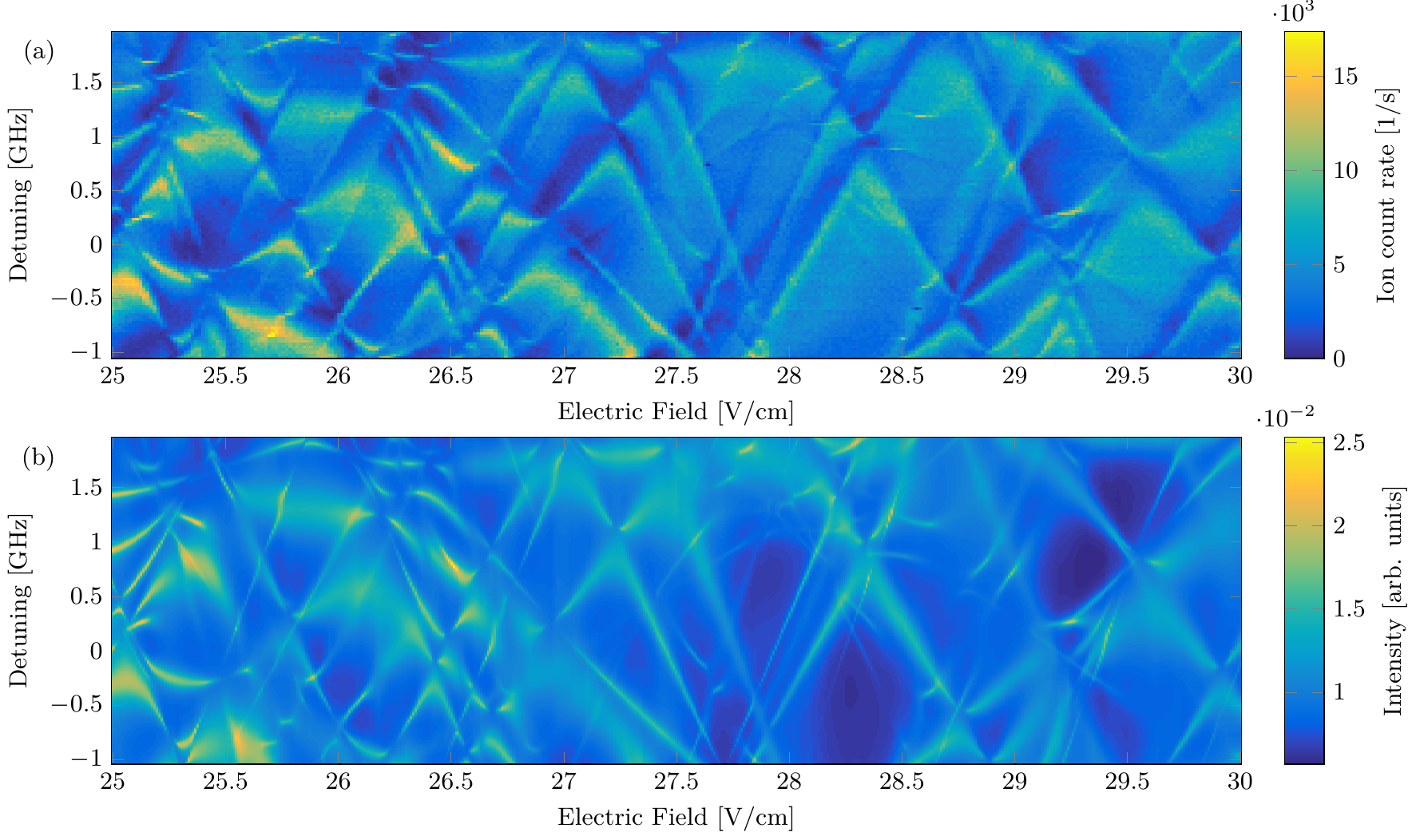}
	\caption{\label{fig:70S-1-2-exp_cap_f_ext_25_30}(Color online) Ionization Stark spectrum near $70\text{S}_{1/2}$. Continuation from Fig. \ref{fig:70S-1-2-exp_cap_f_ext_20_25}.} 
\end{figure*} 

\begin{figure*}
	\includegraphics{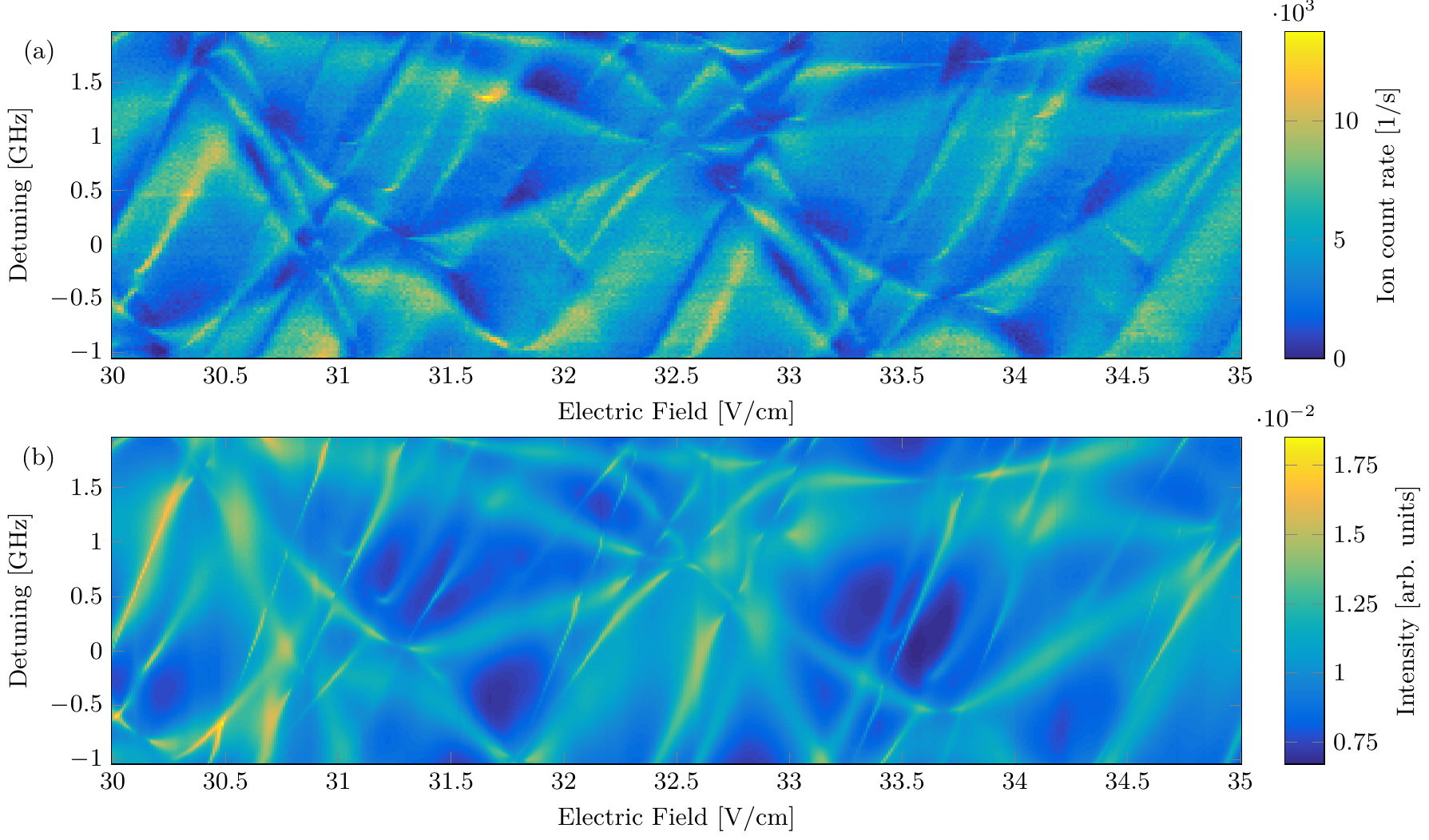}
	\caption{\label{fig:70S-1-2-exp_cap_f_ext_30_35}(Color online) Ionization Stark spectrum near $70\text{S}_{1/2}$. Continuation from Fig. \ref{fig:70S-1-2-exp_cap_f_ext_25_30}.} 
\end{figure*} 

\begin{figure*}
	\includegraphics{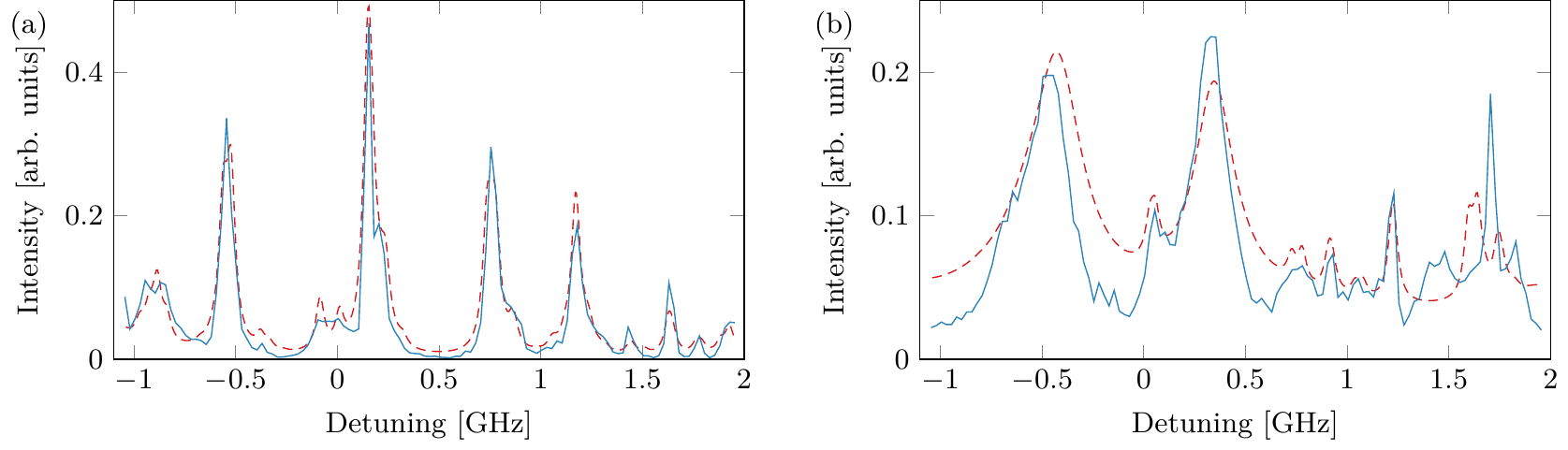}
	\caption{\label{fig:70S-1-2_freqlines}(Color online) Ionization Stark spectra from the experiment (blue solid lines) and the numerical calculations (red dashed lines) near $70\text{S}_{1/2}$ at (a) \SI{16.53}{\V\per\cm} (see Fig. \ref{fig:70S-1-2-exp_cap_f_ext_15_20}) and (b) \SI{24.10}{\V\per\cm} (see Fig. \ref{fig:70S-1-2-exp_cap_f_ext_20_25}). While we find agreement within our experimental resolution of \SI{25}{\MHz} in (a), some small deviations can be seen in (b). The two highest peaks from the experimental data in (b) have a slightly asymmetric line shape, which may indicate possible Fano resonances. Furthermore, we find discrepancies around \SIrange{1.3}{2}{\GHz}, which we attribute to drifts of the voltage source that was used in the experiment and the high sensitivity of these resonances to the electric field.} 
\end{figure*} 

We present experimental results in comparison to results from our numerical calculations near the unperturbed state $43\text{S}_{1/2}$ in Fig. \ref{fig:43S-1-2-exp_cap}. The experimental resolution of the voltage was \SI{100}{\mV}, equivalent to \SI{30.6}{\mV\per\cm}, and the step size of the electric field in the numerical calculations was \SI{40}{\mV\per\cm}. The classical ionization threshold for this state is at \SI{\approx127}{\V\per\cm}, so it is located just within the lower limit of the electric-field range of this figure. The variation of the free parameter $\eta$ described in Sec. \ref{sec:th} led to a value of $\eta=\num{1e6}$ in the depicted region and we used \num{\approx4000} states for the matrix representation of $\hat{H}_\mathrm{CAP}$. The results from the experiment and numerical calculations are in very good agreement. Both data sets clearly show a general broadening of the lines with increasing electric field as one would naively expect above the classical ionization threshold. However, we also find some resonances which are still narrow even at high electric fields as well as overlapping narrow and broad resonances, for example at \SI{\approx143}{\V\per\cm} in Figs. \ref{fig:43S-1-2-exp_cap}(a) and \ref{fig:43S-1-2-exp_cap}(b). 

The results near the unperturbed state $70\text{S}_{1/2}$ are depicted in Figs. \ref{fig:70S-1-2-exp_cap_f_ext_15_20}-\ref{fig:70S-1-2-exp_cap_f_ext_30_35}. The experimental results were recorded with a resolution of \SI{50}{\mV}, equivalent to \SI{15.4}{\mV\per\cm}, and the numerical calculation was performed using steps of \SI{10}{\mV\per\cm}. The ionization threshold near this state is within the range of Fig. \ref{fig:70S-1-2-exp_cap_f_ext_15_20} at \SI{\approx16.1}{\V\per\cm}. In this region, the variation of the free parameter $\eta$ resulted in a value of $\eta=\num{2e5}$ and the matrix for $\hat{H}_\mathrm{CAP}$ is represented using \num{\approx10000} states. Again, we find a general broadening of the resonances beyond the classical ionization threshold. This broadening increases as the electric field gets stronger, but even in the high-field range of Fig. \ref{fig:70S-1-2-exp_cap_f_ext_25_30} we still find some narrow resonances with linewidths on the order of magnitude of our experimental resolution of \SI{25}{\MHz}. 

In the highest-field range (see Figs. \ref{fig:70S-1-2-exp_cap_f_ext_25_30} and \ref{fig:70S-1-2-exp_cap_f_ext_30_35}), the larger ionization rates result in overlapping resonances with different linewidths. In some parts of these spectra, the results from the numerical calculations deviate from the experimental results showing anti-resonance-like features instead of resonances, e.g., in Fig. \ref{fig:70S-1-2-exp_cap_f_ext_30_35}(a), where such a feature starts at \SI{30.5}{\V\per\cm} near \SI{-1}{\GHz} and crosses the whole frequency range of the recorded spectrum up to \SI{31}{\V\per\cm} and \SI{2}{\GHz}. These parts of the spectra, where we find deviations from Lorentzian line shapes, may be interpreted as Fano resonances \cite{Fano.1961}. A Fano-type resonance arises due to interference between different excitation paths to an ionized continuum state. Such Fano resonances have been observed previously in similar systems \cite{Feneuille.1979,Luk.1981,Harmin.1982}. In the present case, besides the discrete Rydberg state that decays to a continuum state by autoionization, there is also the possibility of direct photoionization. Furthermore, several overlapping Rydberg resonances with possibly different ionization rates may lead to more complicated interference effects and thereby more convoluted spectral features. A complete study of these possible Fano resonances may yield a more accurate prediction of the spectrum, but is beyond the scope of this work. 

The numbers of states that were used for the matrix representations of the Hamiltonians were determined in order to calculate a dataset which is bigger than what is shown in this work. We estimate that \num{\approx3000} and \num{\approx6000} states should be enough to reproduce our results near $43\text{S}_{1/2}$ and $70\text{S}_{1/2}$, respectively. Therefore, all results from the numerical calculations are well converged in the regions depicted in Figs. \ref{fig:43S-1-2-exp_cap}-\ref{fig:70S-1-2-exp_cap_f_ext_30_35}. Some discrepancies between the results may arise from slow drifts of the voltage source that was used in the experiment. Since the measurement was performed for one horizontal line in the data after the other, this results in a horizontal mismatch of the experimental data of up to \SI{\pm 0.05}{\V\per\cm} in comparison to the numerically calculated results. Another minor source of deviations between the results may arise since the detection efficiency in the experiment changes as the extractor voltage is increased. Therefore, the ion counts we obtain from the experiment do not scale directly to the results for the intensity from the numerical simulations on the whole range simultaneously. However, this effect is small for the figures presented here since the color maps have been rescaled for each electric-field range separately. 

For a more detailed analysis of the results from the experiment and the numerical calculations, we present cuts of the spectra at two different values of the electric field from Figs. \ref{fig:70S-1-2-exp_cap_f_ext_15_20} and \ref{fig:70S-1-2-exp_cap_f_ext_20_25} in Fig. \ref{fig:70S-1-2_freqlines}. The data shown in Fig. \ref{fig:70S-1-2_freqlines}(a) corresponds to an external electric field of \SI{16.53}{\V\per\cm}, i.e., just above the classical ionization threshold. In this region, the peak positions and widths of the experimentally detected and numerically calculated ionization spectra agree within our experimental frequency resolution of \SI{25}{\MHz}. For the higher electric field of \SI{24.10}{\V\per\cm} shown in Fig. \ref{fig:70S-1-2_freqlines}(b), we find that the shapes of the two resonances with the highest intensities deviate slightly, namely, by an asymmetry of the experimentally obtained signals. We see these asymmetries as another possible manifestation of Fano resonances in this system. Further deviations in the region between \num{1.3} and \SI{2}{\GHz} can be mapped to an electric-field region of \SI{\pm 0.05}{\V\per\cm} and are caused by drifts of the voltage source, as discussed before, in combination with the high sensitivity of these particular resonances to the external electric field (see Fig. \ref{fig:70S-1-2-exp_cap_f_ext_20_25}).

\section{\label{sec:conclusion}Conclusion} 

In this work, we have implemented an extension of the CAP method, in which we alter the shape of the complex absorbing potential according to the change of the external electric field. Furthermore, we have presented experimental results from a setup in which rubidium atoms are continuously ionized through Rydberg states in the presence of an external electric field. Our experimental data show the existence of rich ionization spectra with sharp resonances even far beyond the classical ionization threshold. The presented numerical calculations are capable of predicting the measured spectra of highly Stark-shifted Rydberg states, including the sharp resonances. Thus it is suitable for improving control over the excitation to such states and the subsequent ionization process. An in-depth theoretical study of the possible Fano resonances that we have observed in the experimental data may improve this control even further.

\appendix* 

\begin{acknowledgments}
	This work was financially supported by the FET-Open Xtrack Project HAIRS and by Deutsche Forschungsgemeinschaft through SFB TRR21 and SPP 1929 (GiRyd). M.S. acknowledges financial support from Landesgraduiertenf\"orderung Baden-W\"urttemberg. We thank Peter Schmelcher for advice on complex rotation methods. We thank Nils Schopohl and N\'{o}ra S\'{a}ndor for helpful discussions. 
\end{acknowledgments}

\bibliography{Literatur}

\end{document}